\documentclass[times ,twocolumn, final]{elsarticle}
\usepackage{amssymb}
\usepackage{latexsym}

\usepackage{url}
\usepackage{xcolor}

\usepackage{amsmath} 
\usepackage{CJKutf8}
\usepackage{gensymb}

\usepackage[colorlinks=true, linkcolor=blue, anchorcolor=cyan, urlcolor=blue,citecolor=black]{hyperref}

\definecolor{newcolor}{rgb}{.8,.349,.1}

\begin{document}
\begin{CJK}{UTF8}{gbsn}

\begin{frontmatter}

\title{Plug-and-Play Latent Feature Editing for Orientation-Adaptive Quantitative Susceptibility Mapping Neural Networks}%

\author[1]{Yang Gao \corref{cor1}}
\author[2]{Zhuang Xiong}
\author[3]{Shanshan Shan}
\author[4]{Yin Liu}
\author[4]{Pengfei Rong}
\author[1]{Min Li}
\author[5]{Alan H Wilman}
\author[6]{G. Bruce Pike}
\author[2]{Feng Liu}
\author[2,7]{Hongfu Sun}

\cortext[cor1]{Correspondence: Dr. Yang Gao (Email: \href{yang.gao@csu.edu.cn}{yang.gao@csu.edu.cn}), this manuscript has been accepted by Medical Image Analysis, \href{https://doi.org/10.1016/j.media.2024.103160}{https://doi.org/10.1016/j.media.2024.103160}}

\address[1]{School of Computer Science and Engineering, Central South University, Changsha, China}
\address[2]{School of Electrical Engineering and Computer Science, University of Queensland, Brisbane, Australia }
\address[3]{State Key Laboratory of Radiation, Medicine and Protection, School for Radiological and Interdisciplinary Sciences (RAD-X), Collaborative Innovation Center of Radiation Medicine of Jiangsu Higher Education Institutions, Soochow University, Suzhou, China}
\address[4]{Department of Radiology, The Third Xiangya Hospital, Central South University, Changsha, China}
\address[5]{Department of Radiology and Diagnostic Imaging, University of Alberta, Edmonton, Canada}
\address[6]{Departments of Radiology and Clinical Neurosciences, Hotchkiss Brain Institute, University of Calgary, Calgary, Canada}
\address[7]{School of Engineering, University of Newcastle, Newcastle, Australia}

% \received{xx xx 2023}
% \finalform{xx xx 2023}
% \accepted{xx xx 20233}
% \availableonline{xx xx 2023}
% \communicated{xxx}

\begin{abstract}
%%%
Quantitative susceptibility mapping (QSM) is a post-processing technique for deriving tissue magnetic susceptibility distribution from MRI phase measurements. Deep learning (DL) algorithms hold great potential for solving the ill-posed QSM reconstruction problem. However, a significant challenge facing current DL-QSM approaches is their limited adaptability to magnetic dipole field orientation variations during training and testing. In this work, we propose a novel Orientation-Adaptive Latent Feature Editing (OA-LFE) module to learn the encoding of acquisition orientation vectors and seamlessly integrate them into the latent features of deep networks. Importantly, it can be directly Plug-and-Play (PnP) into various existing DL-QSM architectures, enabling reconstructions of QSM from arbitrary magnetic dipole orientations. Its effectiveness is demonstrated by combining the OA-LFE module into our previously proposed phase-to-susceptibility single-step instant QSM (iQSM) network, which was initially tailored for pure-axial acquisitions. The proposed OA-LFE-empowered iQSM, which we refer to as iQSM+, is trained in a simulated-supervised manner on a specially-designed simulation brain dataset. Comprehensive experiments are conducted on simulated and \textit{in vivo} human brain datasets, encompassing subjects ranging from healthy individuals to those with pathological conditions. These experiments involve various MRI platforms (3T and 7T) and aim to compare our proposed iQSM+ against several established QSM reconstruction frameworks, including the original iQSM. The iQSM+ yields QSM images with significantly improved accuracies and mitigates artifacts, surpassing other state-of-the-art DL-QSM algorithms. The PnP OA-LFE module's versatility was further demonstrated by its successful application to xQSM, a distinct DL-QSM network for dipole inversion. In conclusion, this work introduces a new DL paradigm, allowing researchers to develop innovative QSM methods without requiring a complete overhaul of their existing architectures. 

%%%%
\end{abstract}

\begin{keyword}
%% MSC codes here, in the form: \MSC code \sep code
%% or \MSC[2008] code \sep code (2000 is the default)
 Quantitative Susceptibility Mapping (QSM) \sep Magnetic Dipole Field Orientation \sep Plug-and-Play Orientation-Adaptive Latent Feature Editing (PnP OA-LFE) \sep iQSM+
\end{keyword}

\end{frontmatter}

%\linenumbers

%% main text
\section{Introduction}
\label{sec:introduction}

Quantitative susceptibility mapping (QSM) is a novel post-processing technique to extract the tissue magnetic susceptibility distribution from MRI phase data \citep{b1,b2}. It is a valuable tool for studying various brain diseases, including Parkinson’s Disease \citep{b3,b4}, Alzheimer’s Disease \citep{b5,b6}, Multiple Sclerosis \citep{b7,b8,b9}, and intracranial hemorrhage \citep{b10,b11,b12}, as well as healthy brain aging \citep{new1,new2} and brain oxygen levels \citep{new3, new4}. However, the conventional reconstruction of QSM involves multiple non-trivial intermediate steps \citep{b1,b2}, e.g., phase unwrapping, background field removal, and field-to-source dipole inversion, which not only increase the reconstruction time but accumulate errors \citep{b13}.

Different algorithms have been developed for QSM reconstruction. For example, traditional algorithms include best-path \citep{b14} and Laplacian Unwrapping methods \citep{b15} for phase unwrapping, PDF \citep{b16}, LBV \citep{b17}, SHARP \citep{b18}, and RESHARP \citep{b19} methods for background removal, as well as TKD \citep{b20}, iLSQR \citep{b21}, MEDI \citep{b22,b23}, SFCR \citep{b24}, FANSI \citep{new5}, QSMART \citep{new7} and STAR-QSM \citep{b25} for dipole inversion. There are also single-step TGV-QSM methods \citep{b26,b27} and total field inversion methods \citep{b28,b29}. However, these methods all suffer from various drawbacks, e.g., expensive computational load, manual hyper-parameter tuning, and large artifacts near hemorrhage sources \citep{b13,b30}.

Deep learning (DL) methods are emerging as alternatives to traditional methods \citep{b31}, with most focusing on the dipole inversion step \citep{b32,b33,b34,b35,b36,b37,b39,b40,b48,b49,DIPQSM}, resulting in significantly better results than traditional iterative algorithms in a much shorter reconstruction time. QSMnet \citep{b32} first proposed to train a U-net \citep{b47} for QSM dipole inversion by using the \textit{in vivo} local field maps as training inputs, and COSMOS QSM \citep{b38} images as training labels. This training scheme (learning the mapping from \textit{in vivo} field maps to QSM labels reconstructed using traditional algorithms) was further improved by QSMnet+ \citep{b36}, QSMGAN \citep{b39}, VaNDI \citep{b40}, LPCNN \citep{b37}, and MoDL-QSM \citep{b35} via training data augmentation and more advanced model designs (e.g., GAN \citep{b41,b42} or unrolled networks \citep{b43,b44,b45}). AutoQSM \citep{b46} also successfully trained a U-net for QSM reconstructions from total field maps within this scheme. Alternatively, DeepQSM \citep{b33} and xQSM \citep{b34} chose to train deep neural networks (i.e., U-net and Octave U-net) based on purely synthetic data composed of simple geometric shapes and simulated human brain data using the dipole forward model \citep{b18}, respectively, where the training inputs and labels satisfy the underlying dipole convolution model. One method \citep{b49} used adaptive instance normalization to allow resolution-agnostic reconstruction, while AFTER-QSM \citep{b48}  proposed to handle the local field from oblique and anisotropic acquisitions by adding affine transformations into the reconstruction pipelines. There were also some methods designed for phase unwrapping, e.g., PHU-Net \citep{b50}, and background removal, e.g., SHARQnet \citep{b51} and BFRNet \citep{b52}. In addition to these DL methods, which were designed for a certain step in the QSM reconstruction pipeline, our recently proposed iQSM \citep{b13} method was the first deep learning-based single-step QSM reconstruction technique. It enables direct QSM reconstruction from wrapped phase images in an end-to-end network, and outperforms numerous previous state-of-the-art methods through eliminating the error-accumulating intermediate steps.

\begin{figure*}[!t]
\centerline{\includegraphics[width=0.9\linewidth]{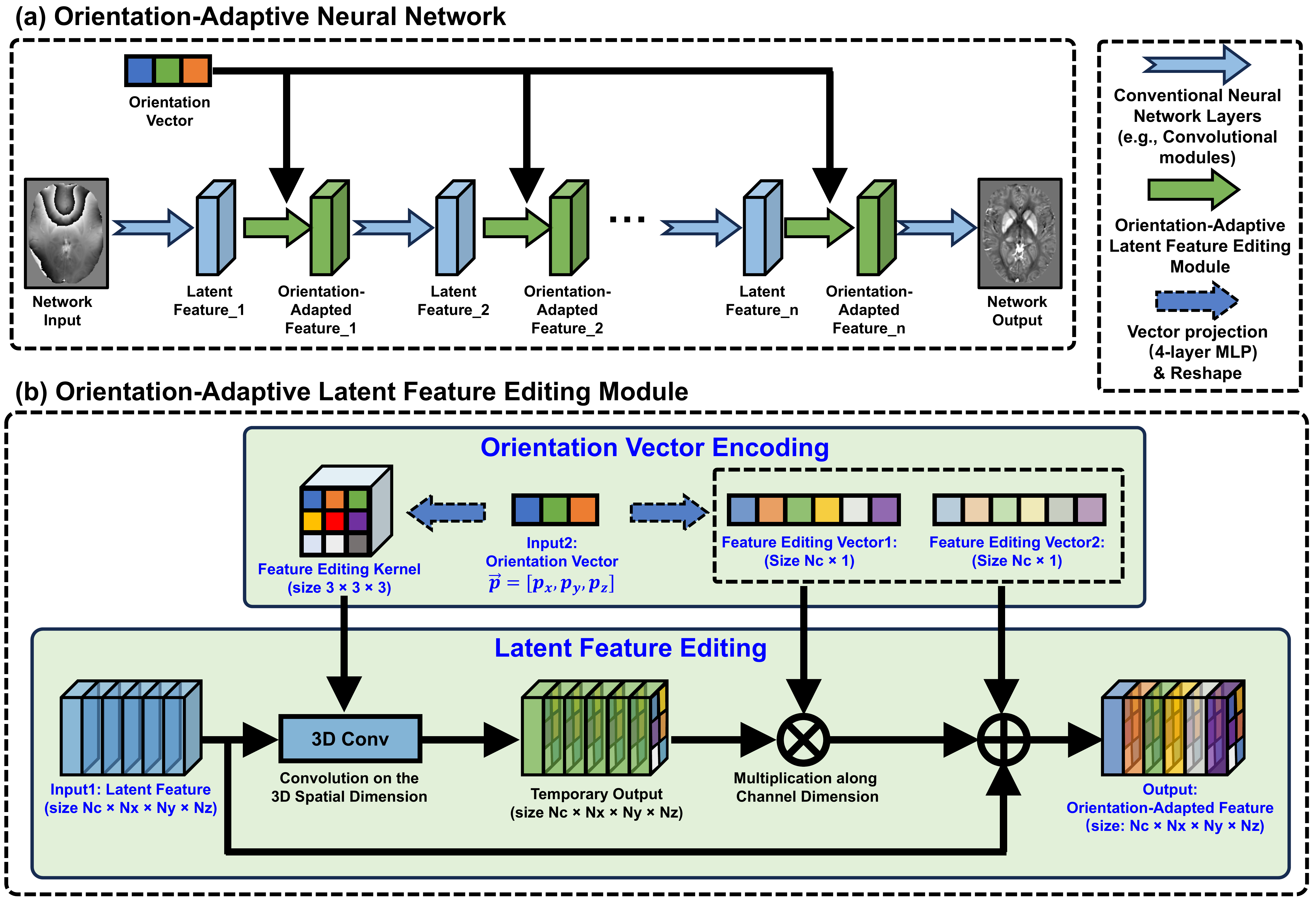}}
\caption{The overall structure of the proposed (a) Orientation-Adaptive Neural Network, which is constructed by incorporating (b) Plug-and-Play Orientation-Adaptive Latent Feature Editing (OA-LFE) blocks onto conventional deep neural networks. The proposed OA-LFE can learn the encoding of acquisition orientation vectors and seamlessly integrate them into the latent features of deep networks.}
\label{fig1}
\end{figure*}

However, the performances of most previous DL-QSM approaches will dramatically decrease when the acquisition orientation with regard to the main magnetic field (or the magnetic dipole field orientation) of the testing data is different from those used in training, which has been reported as a major challenge for current DL-QSM methods in a recent review \citep{b31}. Physics-guided modules can be introduced into DL-QSM methods to enhance the model generalizability to acquisition orientations. Two types of algorithms, i.e., (i) physics-guided unrolled neural networks, e.g., LPCNN \citep{b37} and MoDL-QSM \citep{b35}, and (ii) affine transformation-enabled end-to-end networks, i.e., AFTER-QSM \citep{b48} have demonstrated improvements over conventional pure-network-based QSM solutions. However, the performance of unrolled networks is significantly degraded when applied to data acquired at relatively large oblique angles, and AFTER-QSM is very memory-consuming and sometimes will over-sharpen the QSM images due to a second refinement network design. Moreover, all these methods only perform the dipole inversion process starting from pre-processed local field maps. 

In this work, we propose a novel Orientation-Adaptive Latent Feature Editing (OA-LFE) module, which is designed as Plug-and-Play (PnP) blocks and can be seamlessly embedded into various existing deep neural network backbones, enabling QSM reconstruction from MRI phase measurements at arbitrary acquisition orientations. The proposed PnP OA-LFE is designed to learn the encoding of the acquisition orientation vectors using multi-layer perceptrons (MLPs) and integrate them into the latent features in deep neural networks through 3D image convolution, multiplication, and addition operations. The effectiveness of the proposed OA-LFE module is validated by plugging into our recently developed iQSM network \citep{b13}, which was initially designed for single-step QSM reconstruction at pure-axial acquisitions. This new OA-LFE-empowered network, which we refer to as iQSM+, is trained in a simulated-supervised manner on a specially-designed simulation brain dataset, aiming to directly restore QSM images from MRI raw phase measurements at arbitrary orientations. Extensive simulation and \textit{in vivo} human brain experiments are conducted to compare iQSM+ with multiple established QSM reconstruction pipelines. The OA-LFE module is also plugged into the xQSM neural network, which only performs the dipole inversion task as most of other DL-QSM methods, to demonstrate its universal applicability beyond our chosen base iQSM network. 

\begin{figure*}[!t]
\centerline{\includegraphics[width=0.95\linewidth]{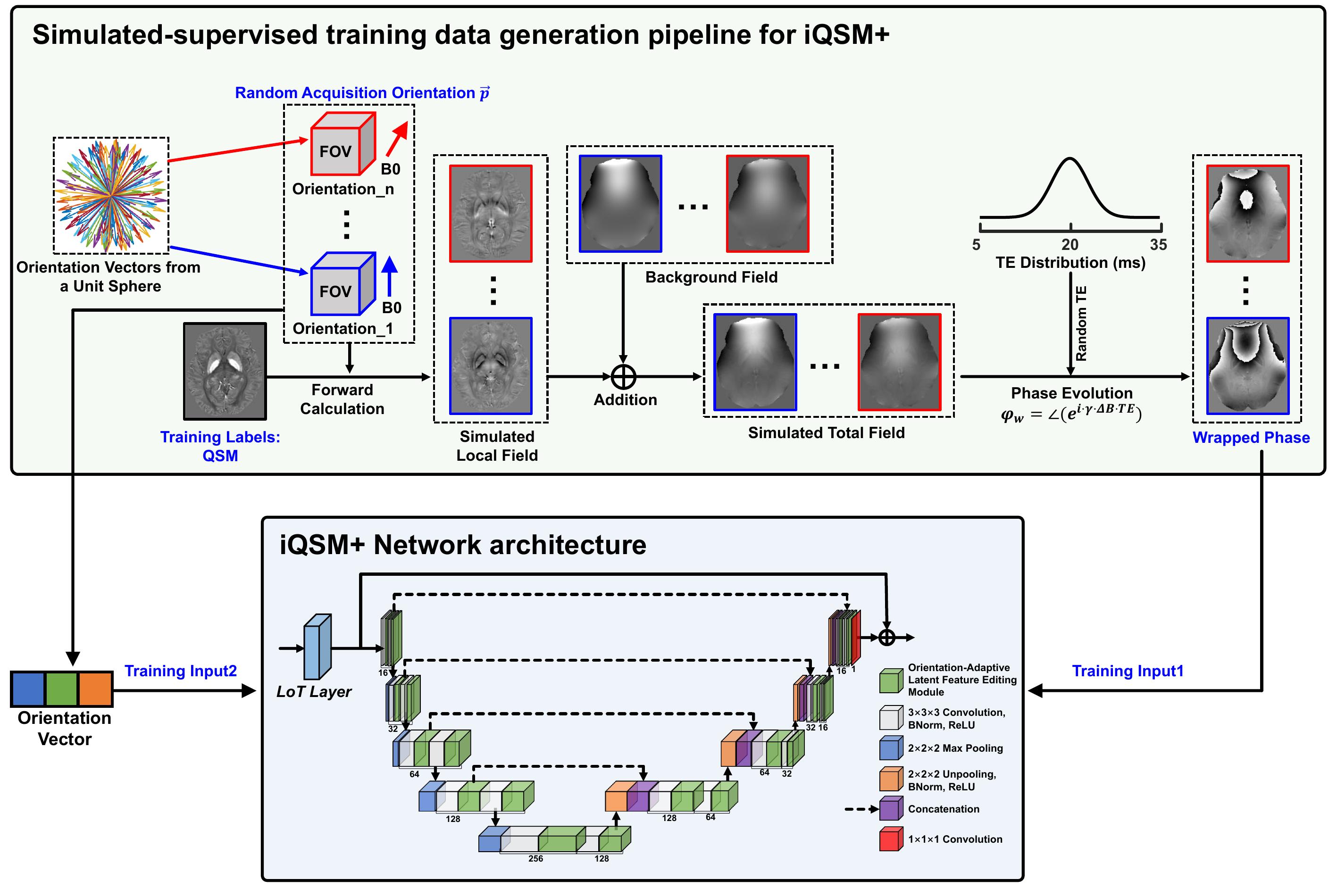}}
\caption{
The proposed iQSM+ network architecture and the associated simulated-supervised training data generation pipeline, including a random dipole orientation vector generation, a forward field calculation for simulating the local field, a background field addition, and a final phase evolution to obtain wrapped phases.}
\label{fig2}
\end{figure*}

\section{Theory and Method}
\label{sec:method}
In this section, we will first demonstrate that single-step QSM reconstruction from MRI raw phases of arbitrary acquisition orientations is an orientation vector-dependent inverse problem. Then, we will introduce the proposed OA-LFE  module and its integration with iQSM to form the new iQSM+ in detail.  

\subsection{Single-step QSM Reconstruction as an Inverse Problem}
The local field ${\delta B}_{local}$ induced by brain tissue is equal to the convolution between the magnetic susceptibility distribution $\chi$ and the unit dipole kernel $D$ in the image domain, which can be expressed as a multiplication in the k-space \citep{new6}: 

\begin{equation}
{\delta B}_{local}(\vec{k})=D(\vec{k})\cdot\chi(\vec{k})
\label{eq1}
\end{equation}
where $D(\vec{k})$ is the unit dipole kernel in k-space,

\begin{equation}
D(\vec{k}) =\frac{1}{3} -\frac{( p_{x} k_{x} +p_{y} k_{y} +p_{z} k_{z})^{2}}{k_{x}^{2} +k_{y}^{2} +k_{z}^{2}}
\label{eq2}
\end{equation}
$\vec{k}$=[$k_{x}$, $k_{y}$, $k_{z}$] are the k-space coordinates; $\vec{p}$=[$p_{x}$, $p_{y}$, $p_{z}$] represents the acquisition (magnetic dipole) orientation unit vector, whose elements are the vector projections of the acquisition Field-of-View (FOV) coordinates onto the main magnetic field ($\vec{B_0}$) direction. In most previous DL-QSM works, only the pure-axial acquisition (i.e., $\vec{p}$=[0, 0, 1]) is considered, which simplifies Eq. \ref{eq2} into Eq. \ref{eq3}: 

\begin{equation}
D(\vec{k}) =\frac{1}{3} -\frac{k_{z}^{2}}{k_{x}^{2} +k_{y}^{2} +k_{z}^{2}}
\label{eq3}
\end{equation}

The relationship between the unwrapped MRI phase and the induced local field perturbation by the tissues inside the brain can be expressed through the Laplacian operator as follows \citep{b15}: 

\begin{equation}
\nabla^{2} \varphi_{u} =2\pi \gamma \cdot B_{0} \cdot TE\cdot \nabla ^{2} \delta B_{local}
\label{eq4}
\end{equation}
where $\varphi_{u}$  is the unwrapped MRI phase; $B_0$ is the main magnetic field strength; $\gamma$ is the gyromagnetic ratio; \textit{TE} is the echo time. Substituting Eq. \ref{eq1} and Eq. \ref{eq2} into Eq. \ref{eq4}: 

\begin{equation}
\begin{split}
\frac{1}{2\pi \gamma B_{0} TE} \nabla ^{2} \varphi _{u} =\frac{1}{3} \cdot \nabla ^{2} \chi -p_{x}^{2}\frac{\partial ^{2} \chi }{\partial x^{2}} -p_{y}^{2}\frac{\partial ^{2} \chi }{\partial y^{2}} -p_{z}^{2}\frac{\partial ^{2} \chi }{\partial z^{2}}\\
 -2p_{x} p_{y}\frac{\partial ^{2} \chi }{\partial x\partial y} -2p_{x} p_{z}\frac{\partial ^{2} \chi }{\partial x\partial z} -2p_{y} p_{z}\frac{\partial ^{2} \chi }{\partial y\partial z}
\end{split}
\label{eq5}
\end{equation}
where [x,y,z] are the image domain coordinates, and the Laplacian of the unwrapped and wrapped phase $\varphi_{u}$ can be calculated using the Laplacian of trigonometric (LoT) functions \citep{b13,b15,b54} from the wrapped phase $\varphi_{w}$ as: 

\begin{equation}
\begin{split}
\nabla ^{2} \varphi _{u}&=LoT(\varphi_w)\\&=cos( \varphi _{w}) \nabla ^{2} sin( \varphi _{w}) -sin( \varphi _{w}) \nabla ^{2} cos( \varphi _{w}) 
\end{split}
\label{eq6}
\end{equation}

Summarizing all above equations, single-step QSM reconstruction from MRI raw phases can be formulated as below, which mathematically describes the single-step QSM reconstruction from MRI wrapped phase images as an orientation-dependent inverse problem since the second-order derivatives can be formulated as linear operators using finite-difference approximations: 

\begin{equation}
A_{\vec{p}}\chi=LoT(\varphi_w)
\label{eq7}
\end{equation}
where $A_{\vec{p}}$ is the orientation-dependent system operator, which is equivalent to the operation of $\left(\frac{1}{3} -p_{x}^{2}\right)\frac{\partial ^{2}}{\partial x^{2}} +\left(\frac{1}{3} -p_{y}^{2}\right)\frac{\partial ^{2}}{\partial y^{2}} +\left(\frac{1}{3} -p_{z}^{2}\right) -2p_{x} p_{y}\frac{\partial ^{2}}{\partial x\partial y} -2p_{x} p_{z}\frac{\partial ^{2}}{\partial x\partial z} -2p_{y} p_{z}\frac{\partial ^{2}}{\partial y\partial z}$. 

In our previous iQSM work \citep{b13}, we have successfully solved Eq. \ref{eq7} at the special case of $\vec{p}$=[0,0,1] (i.e., pure-axial dipole orientation). In this work, a DL-based solution of Eq. \ref{eq7} at arbitrary $\vec{p}$ is proposed below, which can effectively encode and fuse the orientation vector ($\vec{p}$) information into the proposed deep neural networks and learns the inverse of $A_{\vec{p}}$. 

% \section{The Proposed Method}
\subsection{iQSM+ Network Architecture}
As shown in the bottom part of Fig. \ref{fig2}, the network architecture of the proposed iQSM+ is constructed by inserting the proposed novel OA-LFE module into our recently developed iQSM network \citep{b13}, which consists of a learnable LoT Layer and a traditional U-net. The details of the PnP OA-LFE module and the base LoT-Unet from iQSM are detailed as follows: 

\subsubsection{OA-LFE Module} 
As shown in Fig. \ref{fig1}(b), the proposed OA-LFE module takes both the network latent feature as well as the corresponding acquisition orientation vector as inputs, and outputs the Orientation-Adapted Features to the subsequent part of the network. 

\textbf{Orientation Vector Encoding:} Suppose that the input latent feature of OA-LFE module is denoted by $H\in\mathbb{R}^{N_c\times N_x\times N_y\times N_z}$, where $N_c$ is the channel number; $N_x$, $N_y$, and $N_z$ represent the height, width, and depth of the 3D QSM data, respectively. In the LFE module, the orientation vector $\vec{p}$ is first projected into a Feature Editing Kernel (FEK, $K_{fe}\in\mathbb{R}^{3\times3\times3}$), and two Feature Editing Vectors (FEVs, $V_{fe}\in\mathbb{R}^{N_c\times1}$) using three 4-layer MLPs, which can be described as follows: 
% \ref to cite equations. 

\begin{equation}
\begin{array}{l}
Z_{0} =\vec{p}=[p_{x}, p_{y}, p_{z}]^T\\
Z_{n}=SiLU(W_{n} \times Z_{n-1} +b_{n}), \ n = {1, 2, 3}\\
Z_{4} =W_{4} \times Z_{3} +b_{4}
\end{array}
\label{eq8}
\end{equation}
where $Z_{4}$ is the output of the MLP; $W_1\in\mathbb{R}^{3\times3}$, $W_2\in\mathbb{R}^{3\times5}$, and $W_3\in\mathbb{R}^{5\times10}\ $ are the weights of the middle layers, $b_1\in\mathbb{R}^{3\times1}$, $b_2\in\mathbb{R}^{5\times1}$, and $b_3\in\mathbb{R}^{10\times1}\ $ are the associated bias terms; $\times$ represents the matrix multiplication; Sigmoid Linear Unit (SiLU) is adopted as the activate function because it has been demonstrated to improve the performance of shallow networks \citep{SiLu}. For FEK learning, $W_4\in\mathbb{R}^{10\times27}$, $b_4\in\mathbb{R}^{27\times1}$; for FEV learning, $W_4\in\mathbb{R}^{10\times N_c}$, $b_4\in\mathbb{R}^{N_c\times1}$. 

\textbf{Latent Feature Editing:} The input hidden feature $H$ is first convoluted with the learned FEK ($K_{fe}$) to fuse orientation information in spatial dimension:  

\begin{equation}
\begin{array}{l}
H_s=H\ast K_{fe}
\end{array}
\label{eq9}
\end{equation}
where * denotes the 3D image convolution, $K_{fe}\in\mathbb{R}^{3\times3\times3}$ is the learned FEK (reshaped from the MLP output); zero-padding is conducted to make sure that the temporary feature $H_s$ is of the same size as $H$.

Then, the channel-dimension feature editing is conducted to take full advantage of the orientation information:

\begin{equation}
\begin{array}{l}
H_s^c={{V1}_{fe}\cdot H}_s+{V2}_{fe}
\end{array}
\label{eq10}
\end{equation}
where ${V1}_{fe}\in\mathbb{R}^{N_c\times1}$ and ${V2}_{fe}\in\mathbb{R}^{N_c\times1}$ are the two FEVs projected from the orientation vector, respectively. 
Finally, a skip connection from the latent feature input $H$ to temporary edited feature $H_s^c\ $ is added, forming a residual block:

\begin{equation}
\begin{array}{l}
H_{OA}=H+H_s^c
\end{array}
\label{eq11}
\end{equation}
where $H_{OA}{\in\mathbb{R}}^{N_c\times N_x\times N_y\times N_z}$ is the final output of the proposed OA-LFE module, i.e., Orientation-Adapted Feature. 

\subsubsection{Orientation-Adaptive iQSM}
As shown in the bottom part of Fig. \ref{fig2}, the network backbone of the proposed iQSM+ is constructed by combining the proposed PnP OA-LFE modules into our previously developed iQSM network \citep{b13}, which is composed of a novel LoT Layer and a traditional U-net. The OA-LFE modules are appended after every 3D convolutional layer in the U-net part to make full use of the orientation information. 

The U-net part contains 18 three-dimensional convolutional layers with kernel size of 3$\times$3$\times$3 and stride size of 1$\times$1$\times$1 and 1 voxel zero padding to maintain the image size; 18 OA-LFE modules following the 3D convolutional layers to gain Orientation Adaptability; 4 max-pooling and 3D transposed convolution layers of kernel size 2$\times$2$\times$2 are introduced in the contracting and expanding path for multiscale learning, respectively; 22 ReLU units are set as the activation function, and a 1$\times$1$\times$1 3D convolutional layer without zero padding is adopted as the output layer. Finally, a skip connection from the first channel of the LoT Layer output to the U-net output is added, forming a residual block \citep{b55}, which helps the training process converge faster. Furthermore, this skip connection can mitigate the vanishing gradient problem during training \citep{b56}. 

\begin{figure*}[!t]
\centerline{\includegraphics[width=0.95\linewidth]{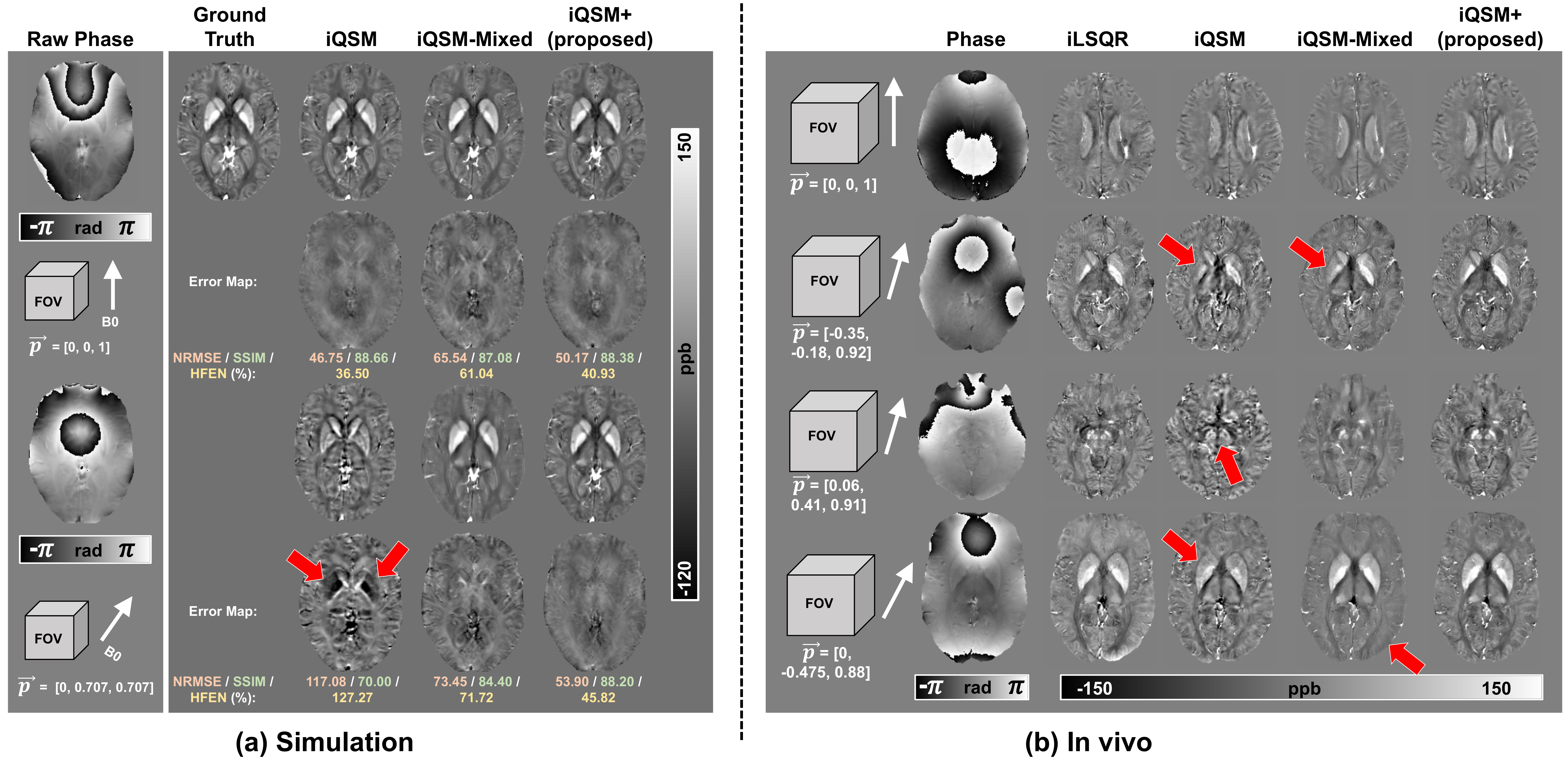}}
\caption{
Comparison of the original iQSM, iQSM-Mixed, and the proposed iQSM+ methods on (a) two simulated brains with different acquisition orientations, and (b) four \textit{in vivo} brains scanned at  multiple 3T MRI platforms. Red arrows point to the reconstruction errors in the original iQSM and iQSM-Mixed.}
\label{fig3}
\end{figure*}

\begin{figure*}[t]
\centerline{\includegraphics[width=0.9\linewidth]{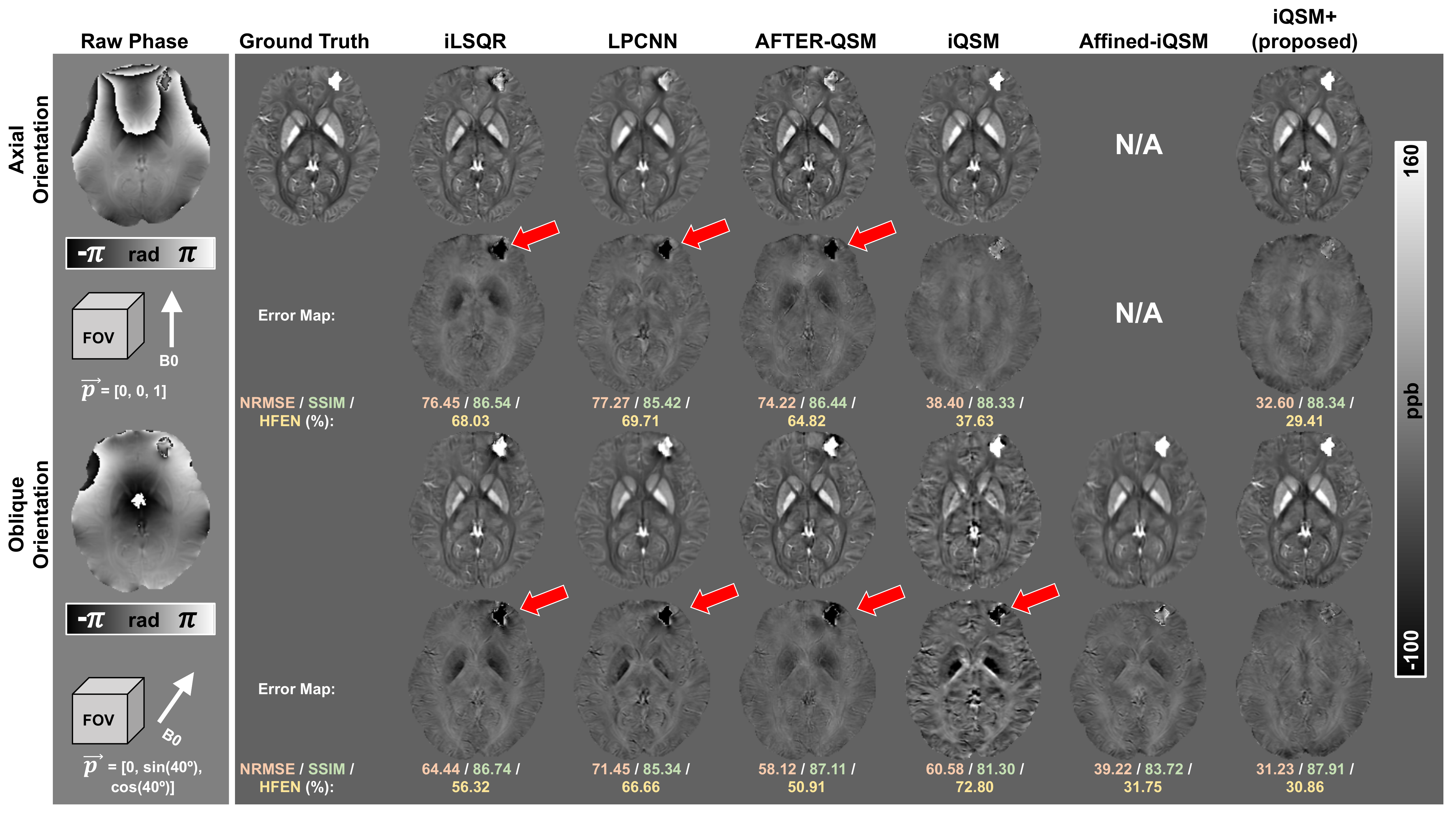}}
\caption{
Comparison of different QSM methods on two simulated pathological brains with a hemorrhage lesion in the frontal white matter. The upper two rows show the results and corresponding error maps relative to the ground truth QSM at neutral head orientation, while the bottom two rows illustrate the results at oblique dipole orientation (40$\degree$  titled). Red arrows point to the apparent errors on the hemorrhage lesion region.}
\label{fig4}
\end{figure*}

\begin{figure*}[t]
\centerline{\includegraphics[width=0.9\linewidth]{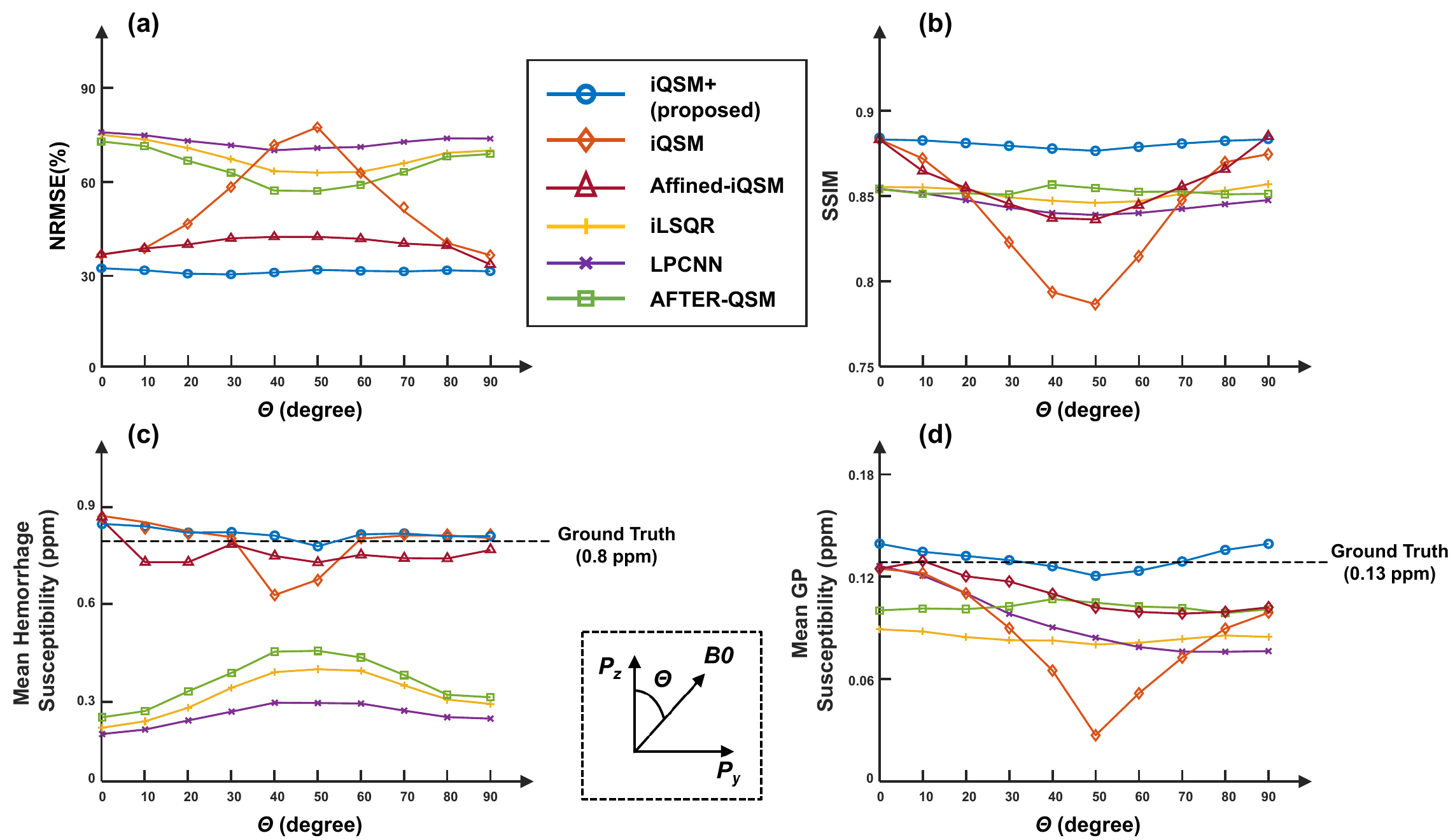}}
\caption{
Analysis of different QSM methods against the data acquisition orientations on 10 phase images simulated from a $\chi_{33}$  label. The line curves of different QSM results on four quantitative metrics, i.e., (a) NRMSE, (b) SSIM, (c) mean hemorrhage susceptibility, and (d) mean Globus Pallidus (GP) susceptibility are compared.}
\label{fig5}
\end{figure*}

\begin{figure*}[!t]

\centerline{\includegraphics[width=0.95\linewidth]{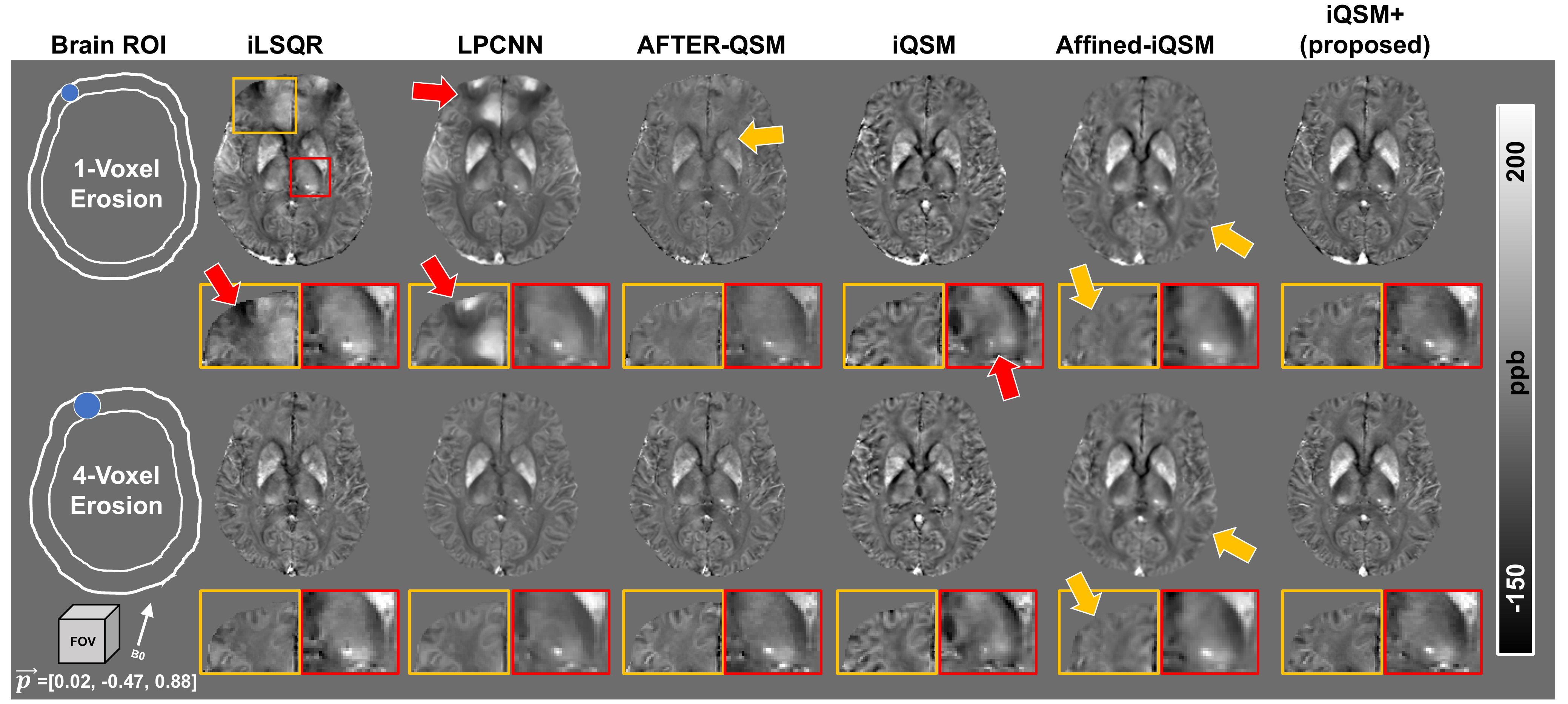}}
\caption{
Comparison of susceptibility maps computed from various QSM frameworks on an \textit{in vivo} SH patient at 3T. The top two rows demonstrate the results with 1-voxel brain edge erosion, while the bottom two rows illustrate the result of 4-voxel mask erosion. The brain edge erosion was performed during the RESHARP background removal step for the LU+RESHARP-based QSM pipelines, while for the iQSM-related approaches, brain edge erosion was applied to the LoT layer outputs before inputting into the U-net part. Red arrows point to the reconstruction errors and artifacts in iQSM, iLSQR, and LPCNN, while yellow arrows point to the apparent over-smoothing and contrast loss in Affined-iQSM and AFTER-QSM.}
\label{fig6}
\end{figure*}

\section{Experiments}
\label{sec:experiments}
In this section, we first describe the details concerning the network training for the proposed iQSM+. Then, we introduce how the comparative experiments were conducted to evaluate the performances of the proposed iQSM+ on MRI phase data acquired from various platforms. Institutional ethics board approvals have been obtained for all MRI brain data used in this work. 

\subsection{Simulated-supervised Training Data Generation}
A simulated-supervised training scheme was introduced to simulate varying orientation vectors spanning the entire 3D unit sphere for network training. As shown in the top part of Fig. \ref{fig2}, unit orientation vectors were first simulated using $\vec{p}=$[sin⁡($\theta$)cos($\phi$), sin⁡($\theta$)sin($\phi$), cos⁡($\theta$)], where $\theta\sim[0,\pi]$ and $\phi\sim[0,2\pi]$ are random spherical coordinates varying from batch to batch. Next, simulated local field maps of various head orientations were first calculated from the QSM ground truths using forward dipole convolution. Then, the corresponding background field maps were added to generate the total field images. Finally, the wrapped phases were simulated from the total field using a previously introduced phase evolution method, scaling with the main field strength and echo times \citep{b13}. The above simulation process from the QSM labels to the training inputs, i.e., the wrapped phases, was encapsulated as a transform function. In this work, only training labels were explicitly prepared before network training, and the corresponding training inputs were dynamically created using the preprocessing transform function at the beginning of each training batch. A major advantage of this simulated-supervised training scheme is that we could simulate as many acquisition orientation vectors as desired with a limited number of training labels. In this study, the proposed iQSM+ were trained based on 14400 training patches for 100 epochs, which indicated that in total 14400$\times$100 head orientation vectors could be simulated using the proposed simulated-supervised training scheme.

The proposed iQSM+ was trained based on 14400 small QSM patches (size: 64$\times$64$\times$64) cropped from 96 full-sized QSM volumes (size: 144$\times$196$\times$128), which were acquired from 96 healthy volunteers at 3T (GE Discovery 750) using a 3D multi-echo GRE sequence (ME-GRE) with the following parameters: First TE/{$\Delta$TE}/TR = 3/3.3/29.8 ms, 8 unipolar echoes, FOV = 144$\times$196$\times$128 mm$^3$, voxel size=1 mm$^3$ isotropic, total scan time = 5.3 min. To obtain the full-sized QSM labels, a traditional QSM reconstruction pipeline, i.e., BET \citep{b57} for brain mask segmentation, Laplacian Unwrapping for phase unwrapping and total field estimation, RESHARP \citep{b19} with 3-voxel brain erosion on BET brain mask for accurate local field calculation, and iLSQR \citep{b21} for dipole inversion, was performed on the 96 full-size ME-GRE raw phase images. We also used the PDF method to estimate the corresponding background field maps for each QSM training label. The cropping was carried out by sliding a 64$\times$64$\times$64 window to traverse the full-size volumes with a stride of 16$\times$22$\times$12. Similar to our recent iQSM work, a total of 14400 simulated pathological QSM patches were also generated using our previously developed simulation pipeline for each healthy QSM patch, and then the two datasets were randomly mixed at the beginning of each training epoch during network training, as described in our recently proposed iQSM training strategies \citep{b13}.  

\subsection{Network Training}
The loss function used for iQSM+ training consists of two parts, i.e., the Mean-Squared-Error (MSE) loss between the network QSM predictions and the QSM training labels, and a model loss to measure the difference between the ground truth local field maps and the estimated local field calculated from the predicted QSM images:  

\begin{equation}
\begin{split}
L\left(\theta\right) &= L_{MSE}\left(\theta\right)+\lambda L_{Model}\left(\theta\right) \\ 
&=\frac{1}{N}\sum_{n=1}^{N}(\left\| y_n- \tilde{y}_n(\theta)) \right\|_{2}^{2} + 
      \lambda\left\| D*y_n- D*\tilde{y}_n(\theta)) \right\|_{2}^{2})
\end{split}
\label{eq12}
\end{equation}
where $\theta$ denotes the network parameters to be optimized, $y_n$ represents the training labels (ground truth QSM data), and $\tilde{y}_n(\theta)$ is the network output QSM prediction, $D$ is the dipole kernel, $n = \{1, 2, 3, \dots, N \}$ is the data index, $N$ is the batch size, and $\lambda$ is a weighting parameter between the two losses. In this work, $\lambda$ is empirically set to 0.1. 

All neural networks in this study were implemented using PyTorch 1.8. The source codes for iQSM+ along with pre-trained network checkpoints are available at \url{https://github.com/sunhongfu/deepMRI/tree/master/iQSM_Plus}. The network parameters were initialized with Gaussian random variables with a mean of zero and a standard deviation of 0.01, except for the convolutional kernels in LoT Layer, which were initialized using the 27-point-stencil Laplacian operators \citep{b13}. The training batch size was set to 32 and the network was trained for 100 epochs using the Adam optimizer. The learning rate was set to $10^{-3}$, $10^{-4}$, and $10^{-5}$ for the first 40, 40-80, and the final 20 epochs, respectively. It took around 24 hours to finish the network optimization on one Nvidia Tesla A6000 GPU. 

\subsection{Evaluation Datasets}
To demonstrate the performance and generalizability of the proposed iQSM+, extensive comparative experiments were evaluated on a large number of simulated and \textit{in vivo} brain datasets from different MRI platforms, including: 
\begin{enumerate}
    \item Simulated GRE phase measurements from two COSMOS healthy brains (dipole orientation vectors $\vec{p}=$[0,0,1] and [0,0.707,0.707]) were tested in an ablation study to quantitatively evaluate the effectiveness of the novel OA-LFE modules. 
    \item Raw phase measurements from 4 healthy volunteers were acquired \textit{in vivo} on different MRI scanners. The first two were scanned using the same acquisition parameters as the training data with acquisition orientations $\vec{p}$=[0,0,1] and [-0.35,-0.18,0.92], respectively. The third data were scanned at 3T (Phillips Ingenia Elition X) with the following parameters: 7 unipolar echoes, First TE/{$\Delta$TE}/TR=4.1/4/40 ms, FOV = 224$\times$224$\times$128 mm$^3$, voxel size=1 mm$^3$ isotropic, acquisition orientation vector $\vec{p}=$[0.06,0.41,0.91], total scan time = 4.2 min. The fourth data was scanned at 7T (Siemens Magnetom) with 4 unipolar echoes, First TE/{$\Delta$TE}/TR = 5/3/15.5 ms, FOV = 192$\times$256$\times$144 mm$^3$, voxel size=1 mm$^3$ isotropic, acquisition orientation vector $\vec{p}=$[0,-0.475,0.88], total scan time = 3.5 min.
    \item Ten simulated pathological brain data (orientation vectors $\vec{p}=$[0, sin⁡($n*\pi$/9), cos⁡($n*\pi$/9)], $n$=\{0, 1, 2, $\dots$, 9\}) with a simulated hemorrhage source (0.8 ppm) added to a STI-$\chi_{33}$ QSM label \citep{b58,b59}. We used this pathological simulation dataset to explore how varying acquisition angles impact different QSM methods.
    \item A patient with subdural hematoma (SH) was scanned at 3T (Phillips, Ingenia Elition X) using a ME-GRE sequence with the following acquisition parameters: 10 unipolar echoes, acquisition orientation vector $\vec{p}=$[0.02,-0.47,0.88], First TE/{$\Delta$TE}/TR = 4.2/4.1/43.4 ms, FOV = 240$\times$240$\times$140 mm$^3$, 1 mm isotropic voxel size, total scan time = 4.6 min. Two more patients (one with multiple micro bleeding (MB) lesions and one with intracranial hemorrhage (ICH)) were also scanned \textit{in vivo} using this 10-echo sequence at 3T with a pure-axial acquisition orientation ($\vec{p}=$[0,0,1]) to investigate the effects of incorrect orientation vectors on iQSM+ reconstructions.
    \item A total of 100 ME-GRE scans (50 at pure-axial orientation and 50 at oblique orientations) from 8 subjects in a public QSM dataset \citep{b59} acquired at 3T (Siemens Prisma) were used to quantitatively compare deep gray matter (DGM) susceptibility measurements of different QSM methods. 
    
\end{enumerate}

\begin{figure*}[t]
\centerline{\includegraphics[width=0.95\linewidth]{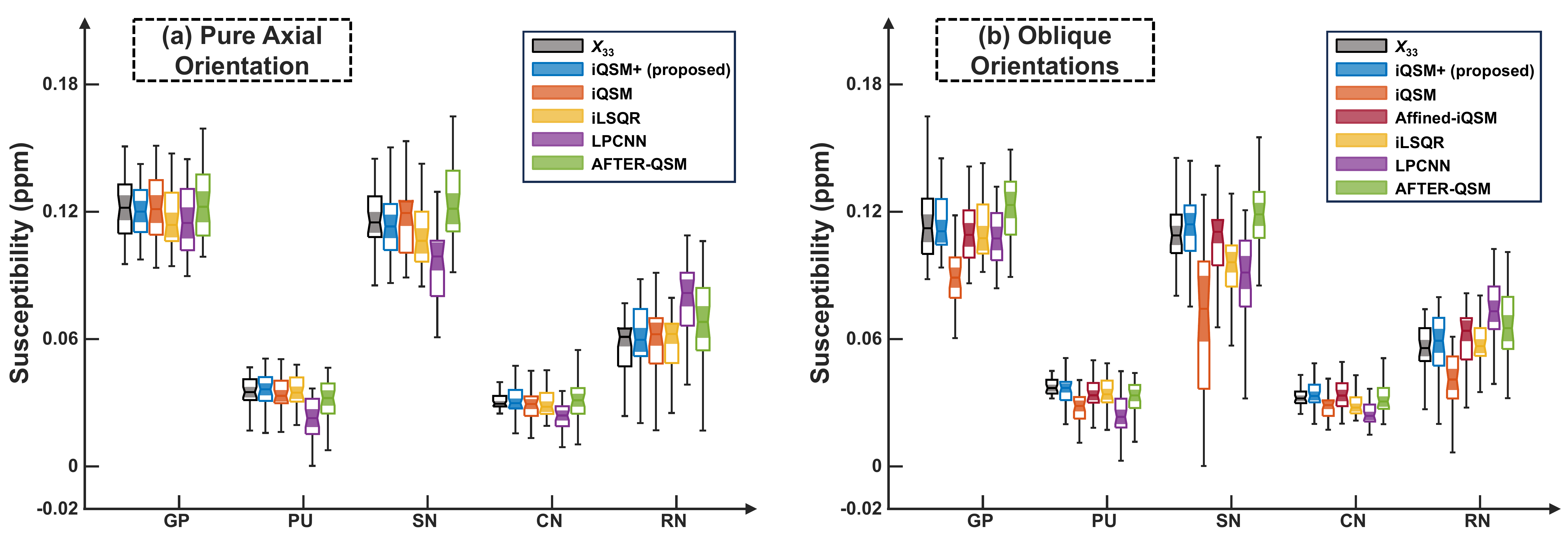}}
\caption{
ROI analysis of different QSM methods on five deep grey matter regions from 100 \textit{in vivo} subjects, categorized into (a) 50 with neutral acquisition orientations, and (b) 50 with oblique orientations.}
\label{fig7}
\end{figure*}

\begin{figure*}[!t]
\centerline{\includegraphics[width=0.95\linewidth]{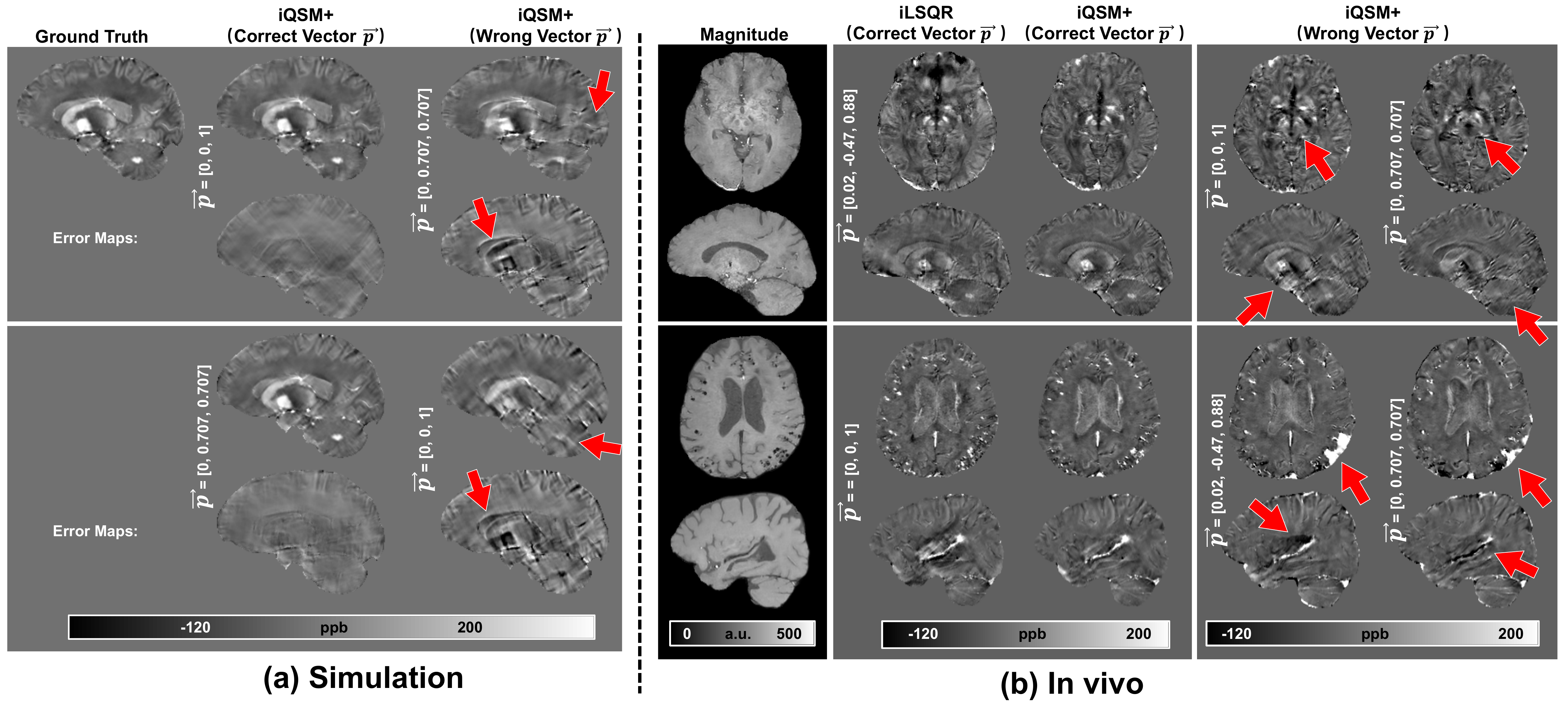}}
\caption{
The effects of incorrect orientation vectors deliberately input to iQSM+ on (a) two simulated brains from the COSMOS brain, and (b) three \textit{in vivo} subjects (i.e., the SH, MB, and ICH patients). The top two rows in (b) show the results from the SH patient, while the bottom two rows in (b) demonstrate the QSM images from the MB and ICH patients, respectively. Red arrows point to the apparent reconstruction errors due to incorrect orientations provided to the network.}
\label{fig8}
\end{figure*}

\begin{figure}[!t]

\centerline{\includegraphics[width=0.95\linewidth]{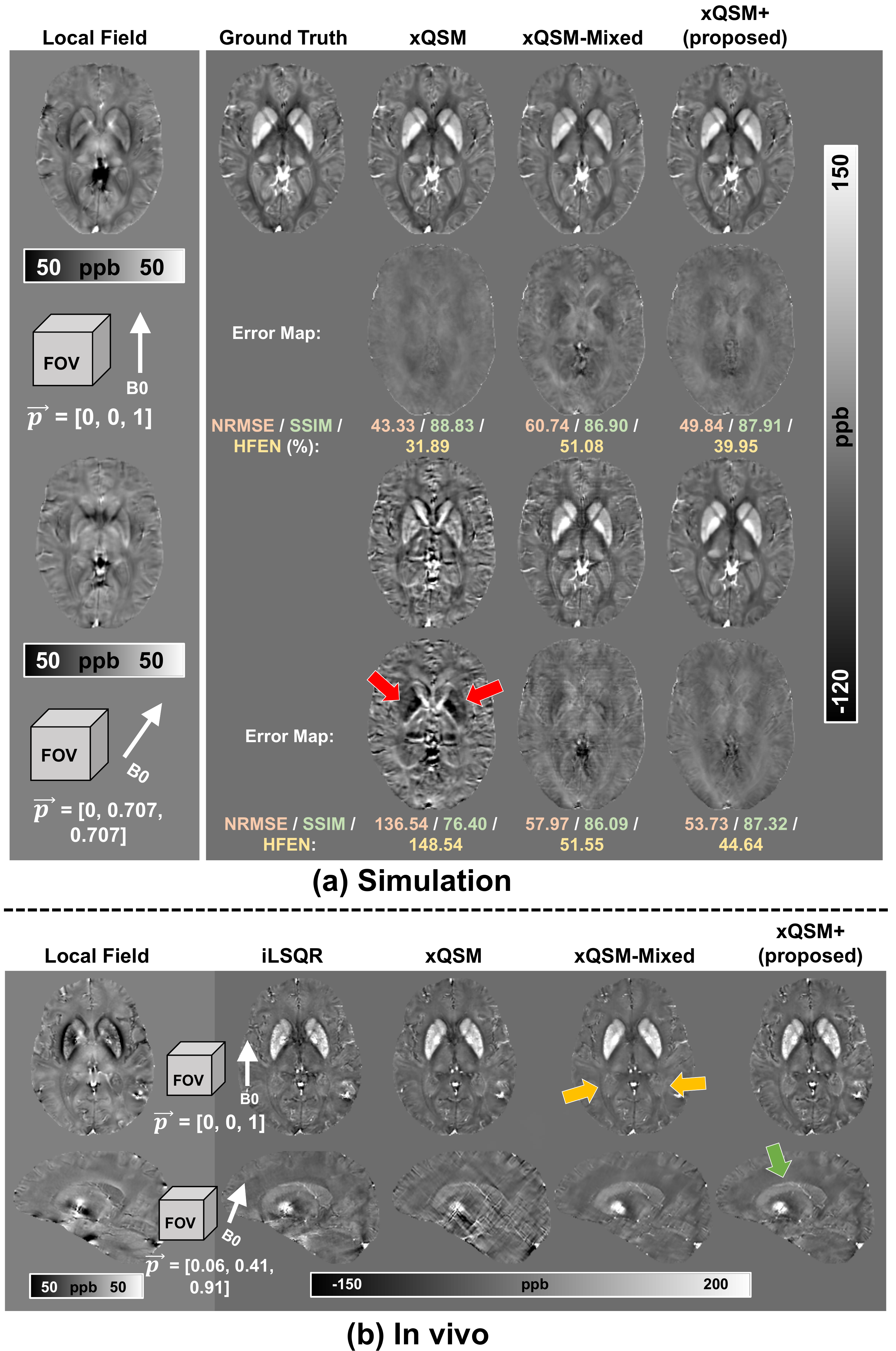}}
\caption{
Comparison of the original xQSM, xQSM-Mixed, and the proposed xQSM+ on (a) two simulated brains from the COSMOS data, and (b) two \textit{in vivo} brain subjects scanned at 3T. Red arrows point to the reconstruction errors in the original xQSM and xQSM-Mixed. The yellow arrows point to the structural loss in xQSM-Mixed, while the green arrow points to the vein that was smoothed out in xQSM-Mixed but successfully preserved in xQSM+.}
\label{fig9}
\end{figure}

\subsection{Performance Evaluation and Numerical Metrics}
To demonstrate the effectiveness of the proposed PnP OA-LFE module, another two iQSM-Nets without the OA-LFE modules, i.e., the original iQSM and iQSM-Mixed, were trained and compared with the proposed iQSM+ on both simulated and \textit{in vivo} QSM data. The original iQSM was trained on phase data of pure-axial orientation, and iQSM-Mixed was trained on the same dataset as iQSM+ with local field maps generated using mixed dipole orientations. Different from iQSM+, the iQSM-Mixed shares the same network architecture as the original iQSM without the proposed OA-LFE module to take in orientation vectors as another input to the network.  

We also compared the proposed iQSM+ with several established multiple-step methods, including iLSQR \citep{b21}, LPCNN \citep{b37}, and AFTER-QSM \citep{b48}, which in theory can handle oblique dipole inversions. All multi-step methods started from Laplacian Unwrapping \citep{b15} and RESHARP \citep{b19} background field removal in this work. Similar to the AFTER-QSM \citep{b48} design, Affined-iQSM was also introduced in this study, i.e., applying affine transformation for image rotation before and after the pretrained iQSM model, where the affine matrix was calculated from the dipole orientation vectors. LPCNN and AFTER-QSM models were trained by their original authors without retraining on our simulated-supervised dataset. All deep learning method inferences were conducted on one Nvidia 4090 GPU, and the traditional non-deep learning algorithms were accomplished on one Intel(R) Core(TM) i7-13700F CPU. 

For quantitative evaluation of different QSM reconstruction frameworks, commonly used numerical metrics \citep{b60} including Normalized Root Mean Squared Error (NRMSE), Structural Similarity (SSIM), and High-Frequency Error Norm (HFEN) were computed and reported. In addition, susceptibility values of five DGM regions, i.e., globus pallidus (GP), putamen (PU), caudate (CN), substantia nigra (SN), and red nucleus (RN) in the one hundred \textit{in vivo} public brain data \citep{b59} were measured and reported as bar graphs with the DGM parcellation provided by their authors. Line graphs on the ten simulated pathological data of the numerical metrics and Region-of-Interest (ROI) (simulated hemorrhage and GP) susceptibility measurements were also comprehensively compared with other QSM methods.

\section{Results}
\label{sec:results}
\subsection{Ablation Study: The Effectiveness of the Proposed PnP OA-LFE Module}

Figure \ref{fig3} compares the proposed iQSM+ with the original iQSM and iQSM-Mixed on two COSMOS-based simulated and four \textit{in vivo} brain subjects. It is clear from the simulation results (Fig. \ref{fig3}(a)) that iQSM+ showed consistent reconstructions for phase data of both pure-axial ($\vec{p}=$[0,0,1]) and 45$\degree$ oblique ($\vec{p}=$[0,0.707,0.707]) orientations, while the original iQSM failed to present reasonable results for the latter case, and iQSM-Mixed demonstrated apparent over-smoothing QSM results. The original iQSM’s NRMSE/SSIM/HFEN performances degraded from 46.75/90.66/36.50$\%$ to 117.08/70.00/127.27$\%$ due to obliqued acquisition, while the maximum HFEN deviations between the two orientations for those of iQSM+ were less than 5$\%$, compared to 91.22$\%$ and 10.68$\%$ for the original iQSM and iQSM-Mixed, respectively. Similar trends can be observed from the \textit{in vivo} results in Fig. \ref{fig3}(b). As indicated by the red arrows, iQSM failed to restore desirable QSM contrast in the GP and SN regions for phase data of oblique acquisition orientations, while iQSM-Mixed presented over-smoothing images. In contrast, the iQSM+ showed consistent QSM results for all four different orientations. This ablation study confirms that the proposed OA-LFE modules significantly enhanced the original iQSM and enabled the new iQSM+ to accurately reconstruct QSM images from arbitrary head orientations. 

An ablation study was also conducted to investigate the effectiveness of the proposed FEV and FEK designs using the 10 simulated pathological brains data as described in Section 3.3 (Evaluation Datasets), with qualitative and quantitative results shown in Supp. Fig. 1 and Supp. Table 1, respectively. The performance of iQSM+ without FEK or FEV module drops significantly, implying both FEV and FEK designs are effective for the QSM reconstruction task, with FEV of higher contribution than FEK. 

Furthermore, the learned FEKs and FEVs in the first OA-LFE module across 10 different acquisition orientations are shown in Supp. Fig. 2. It is observed that FEKs and FEVs are changing continuously with the increasing acquisition orientation angles, which suggests some degree of interpretability and trustworthiness of the OA-LFE models. 

\subsection{Simulated Pathological Dataset}
Figure \ref{fig4} compares the proposed iQSM+ with various established QSM methods on two phase images with a simulated hemorrhage lesion (0.8 ppm) added onto a $\chi_{33}$ brain base. The upper two rows compare different QSM methods on phase data of axial orientation, while the bottom two rows illustrate the results of oblique case. According to the error maps and the numerical metrics, the proposed iQSM+ showed the most consistent and best reconstruction results among all methods. Although Affined-iQSM could help improve the original iQSM’s results of oblique dipole orientations, it smoothed out some fine brain structures. This implies that the proposed OA-LFE modules can better handle orientation vectors than the non-learnable affine transformation based rotations. It is also found that multi-step methods dramatically underestimated the hemorrhage susceptibility comparing to iQSM-basesd single-step methods, which is consistent with the findings in our recent work \citep{b13}. 

More quantitative analyses based on this $\chi_{33}$ label were investigated and reported as line graphs in Fig. \ref{fig5} concerning the robustness of various QSM methods on four evaluating metrics, i.e., NRMSE, SSIM, hemorrhage susceptibility, and GP measurement changes against the dipole orientation vectors (from pure-axial to 90$\degree$ titled). It is clear that the proposed iQSM+ showed the most flat curves along oblique angles, while the original iQSM’s curves changed the most, which further confirmed that the effectiveness of the proposed OA-LFE modules. The proposed iQSM+ displayed the smallest NRMSE and the most accurate hemorrhage measurements. The hemorrhage curves in Fig. \ref{fig5}(c) clearly showed that the multi-step algorithms significantly underestimated hemorrhage susceptibilities.

\subsection{\textit{In vivo} Dataset}
QSM results of various methods on the SH patient were compared in Fig. \ref{fig6}. The upper two rows show reconstructions of 1-voxel brain edge erosion, while the bottom two rows demonstrate results of 4-voxel brain mask erosion. The proposed iQSM+ produced the most visually appealing results for 1-voxel mask erosion data, while iLSQR and LPCNN showed substantial artifacts, AFTER-QSM showed significant contrast loss compared with its 4-voxel erosion result, iQSM failed to produce clear reconstruction of the micro-bleeding, and Affined-iQSM showed noticeable over-smoothing effects. All iQSM-based methods showed more consistent reconstructions between the 1-voxel and 4-voxel mask-eroded data as compared to the first three methods in Fig. \ref{fig6}. 

Box charts in Fig. \ref{fig7} compare different QSM algorithms based on the 100 brain subjects from the public QSM dataset \citep{b59}. The left part plot compares DGM measurements of various methods on data with neutral (pure-axial) orientation, while the right part plot shows the box charts for QSM of oblique orientations. The proposed iQSM+ showed similar measurements as the $\chi_{33}$ reference, traditional iLSQR, and AFTER-QSM for neutral orientation QSM data, while LPCNN showed significant differences in the PU, SN, and RN region. For example, the median values of iQSM+ in PU is 3.9$\%$ deviated from the $\chi_{33}$ reference, 31.40$\%$ lower than LPCNN. Similar trends were found for the oblique subjects, except that the original iQSM trained for pure-axial significantly under-performed all others. 

\subsection{The Influences of Incorrect Orientation Vectors on iQSM+ Results}
The effects of incorrect orientation vectors on iQSM+ are shown in Fig. \ref{fig8}, demonstrating that the proposed OA-LFE indeed learned the effective encoding of dipole orientation vectors instead of simply memorizing 'normal-appearing' QSM results for every possible orientation. According to the error maps in simulation results (Fig. \ref{fig8}(a)), deliberately feeding in incorrect orientation vectors to iQSM+ would result in undesirable reconstruction results with dramatically larger errors. Similar results were also observed in the \textit{in vivo} brain data. As pointed out by the red arrows in Fig. \ref{fig8}(b), providing iQSM+ with the wrong orientation vectors resulted in substantial reconstruction errors and artifacts near the micro bleeds, SN, dental nucleus, and hemorrhage regions.  

\subsection{Applicability of OA-LFE to Other DL-QSM Networks}
All above experiments were carried out to demonstrate the effectiveness and robustness of the proposed OA-LFE for the single-step iQSM. To demonstrate the versatility of the PnP OA-LFE module, we also applied it to another neural network, xQSM \citep{b34}, which was solely designed for dipole inversion starting from the pre-processed local field map. Similar to the ablation studies designed for iQSM+ (Fig. \ref{fig3}), we also trained an xQSM+ (i.e., OA-LFE-empowered xQSM), and compared it with the original xQSM and xQSM-Mixed (similar to the training of iQSM-Mixed), as shown in Fig. \ref{fig9}. Similarly, xQSM+ showed much-improved results with enhanced orientation adaptability than the original xQSM and xQSM-Mixed, confirming that the proposed OA-LFE module is not limited to iQSM, but also effective for QSM dipole inversion neural networks that various research groups have proposed in recent years. 

\section{Discussion and Conclusion}
\label{sec:discussion}
In this work, we propose a novel PnP OA-LFE module  to learn the encodings of dipole orientation vectors and effectively fuse them into deep neural networks, enabling QSM reconstructions from arbitrary acquisition orientations. We developed an iQSM+ method by incorporating the OA-LFE modules into our previously developed single-step iQSM network. To demonstrate the effectiveness of the proposed OA-LFE and the performances of iQSM+, extensive experiments were conducted on simulated and \textit{in vivo} brain subjects from various MRI platforms to compare the proposed method with several established methods including some state-of-the-art DL-QSM methods and the original iQSM. In addition, we also validated the generalizability and versatility of the proposed OA-LFE module on deep neural networks designed for QSM dipole inversion, e.g., xQSM. Overall, this work presents a new DL paradigm enabling QSM researchers to develop novel orientation-adaptive QSM solutions by combining the PnP OA-LFE module into their network structures without a complete overhaul of their existing architectures. 

The proposed iQSM+ was trained using a simulated-supervised framework, which could simulate sufficient dipole orientations needed, without the need to scan a large number of subjects at oblique head orientations. This network training strategy also allowed us to manually control the distribution of the orientation vectors. 

Most previous DL-QSM methods focused on the final QSM dipole inversion step, e.g., QSMnet/QSMnet+ \citep{b32,b36}, xQSM \citep{b34}, LPCNN \citep{b37}, MoDL-QSM \citep{b35} etc., while traditional unwrapping algorithms like Laplacian unwrapping \citep{b15}, and background removal RESHARP \citep{b19} were still necessary as preprocessing steps. In contrast, our previously proposed phase-to-susceptibility iQSM network was able to reconstruct QSM images directly from the MRI raw phases in a single step. Thanks to the PnP property of the proposed OA-LFE, the current version of iQSM+ not only handles arbitrary acquisition orientations, but preserves almost all the advantages of the original iQSM \citep{b13} against the multi-step QSM frameworks, including robustness against brain edge erosion, ultra-fast reconstruction speed, and reduced errors from the intermediate steps, especially phase unwrapping as we have shown in our iQSM work \citep{b13}. 

We evaluated the performance of the proposed iQSM+ on extensive MRI datasets scanned from multiple platforms with various sequence parameters (e.g., TR, TE, flip angle, number of echoes, etc.), which ensured a fair, comprehensive, and realistic comparison for different QSM methods. It is found that the proposed iQSM+ consistently showed improved results on QSM data of different parameters, compared with the original iQSM and other multi-step QSM methods. 

In this study, we limited our experiments of the proposed iQSM+ and xQSM+ to human brain data scanned at 1 mm$^3$ isotropic resolution. Data of anisotropic resolution would require image upsampling to isotropic voxel sizes. Theoretically, retraining or fine-tuning the iQSM+ network with matching anisotropic image resolution will improve the network’s performance; however, this is more computationally expensive and time-consuming. In the future, we will investigate strategies such as those proposed in AFTER-QSM \citep{b48} and AdaIN-QSM \citep{b49} to more effectively and accurately process QSM data of anisotropic resolutions. In addition, the proposed Orientation-Adaptive (OA-LFE module embedded) neural networks were trained in a simulated-supervised manner, where the training labels reconstructed using previously developed QSM methods may contain artifacts not corresponding to brain anatomies and thus influence the network performance. Furthermore, if there is severe subject motion in the MRI scans (e.g., due to respiratory effects or severe head motion), specially tailored motion correction algorithms, including deep learning-based methods, should be performed as preprocessing steps before the iQSM+ reconstruction to eliminate or suppress potential artifacts. Future work may investigate unsupervised algorithms to directly solve QSM from the raw MRI phase acquired at arbitrary acquisition orientations and spatial resolutions. 

\section*{Data and Code Availability Statements}
Data are available on request due to privacy/ethical restrictions. Source codes and trained networks are available at: \url{https://github.com/sunhongfu/deepMRI/tree/master/iQSM_Plus}.

\section*{Acknowledgments}
YG acknowledges that this work was supported by the National Natural Science Foundation of China under Grant No.	62301616, and SS acknowledges support from the National Natural Science Foundation of China under Grant No. 62301352. HS acknowledges support from the Australia Research Council (DE20101297 and DP230101628). GBP acknowledges support from Canadian Institutes of Health Research (FDN-143290).

%%Harvard
\bibliographystyle{model2-names.bst}\biboptions{authoryear}
\bibliography{iQSMPlus}

\end{CJK}
\end{document}